\RequirePackage{fix-cm}
\documentclass[smallextended]{svjour3}       % onecolumn (second format)
\smartqed  % flush right qed marks, e.g. at end of proof
\usepackage{graphicx}
\usepackage{amsmath}
\usepackage{amssymb}
\usepackage{algorithmic}
\usepackage{algorithm}
\usepackage{array}

\begin{document}

\title{Mining Target Attribute Subspace and Set of Target Communities in Large Attributed Networks%\thanks{Grants or other notes
}

\titlerunning{Mining Target Attribute Subspace and Set of Target Communities}        % if too long for running head

\author{Peng Wu         \and
        Li Pan %etc.
}


\institute{Peng Wu \and Li Pan (corresponding author)\at
              School of Electronic Information and Electrical Engineering, Shanghai Jiao Tong University, Shanghai, China. \\
              National Engineering Laboratory for Information Content Analysis Technology, Shanghai Jiao Tong University, Shanghai, China.\\
              \email{catking@sjtu.edu.cn}           %  \\
           \and
           Li Pan \at
              \email{panli@sjtu.edu.cn}
}

\date{Received: date / Accepted: date}

\maketitle

\begin{abstract}
Community detection provides invaluable help for various applications, such as marketing and product recommendation. Traditional community detection methods designed for plain networks may not be able to detect communities with homogeneous attributes inside on attributed networks with attribute information. Most of recent attribute community detection methods may fail to capture the requirements of a specific application and not be able to mine the set of required communities for a specific application. In this paper, we aim to detect the set of target communities in the target subspace which has some focus attributes with large importance weights satisfying the requirements of a specific application. In order to improve the university of the problem, we address the problem in an extreme case where only two sample nodes in any potential target community are provided. A Target Subspace and Communities Mining (TSCM) method is proposed. In TSCM, a sample information extension method is designed to extend the two sample nodes to a set of exemplar nodes from which the target subspace is inferred. Then the set of target communities are located and mined based on the target subspace. Experiments on synthetic datasets demonstrate the effectiveness and efficiency of our method and applications on real-world datasets show its application values.
\end{abstract}

\section{Introduction}
\label{intro}
Community detection \cite{Zhou2009Graph,Xu2012A,Ruan2013Efficient,Moser2009Mining,Silva2012Mining,Gunnemann2013Spectral,Gunnemann2014Gamer,Galbrun2014Overlapping,Huang2015Dense} in networks has attracted a lot of attention for it provides invaluable help for various applications. Many real-world networks are attributed networks where nodes are associated with attributes describing their semantic information. An attributed network consists of nodes, edges and attribute vectors associated with the nodes. Traditional community detection methods \cite{Raghavan2007Near,Wu2015Multi} which consider plain networks without attributes may fail to detect communities with coherent attributes inside. Most of recent attribute community detection methods proposed for attributed networks either treat all available attributes as equally important \cite{Zhou2009Graph,Xu2012A,Ruan2013Efficient}, or take an unsupervised technique to decide the importance weights of attributes \cite{Gunnemann2013Spectral,Gunnemann2014Gamer,Huang2015Dense}. The importance weights of attributes make up an attribute subspace vector for attribute similarity computation. The communities in an subspace are structurally dense and well-separated from the rest of the network, as well as have large similarity under such subspace. A community have large similarity under a subspace when it have similar values on the attributes with large weights in such subspace. For specific application, set of \textit{target communities} in certain \textit{target subspace} rather than all communities are usually required. The essence of application requirement is captured by those attributes with large weights in the target subspace. These attributes are called \textit{focus attributes}.

A toy example is given in Fig. 1. The network represents the friendship relations and each node is associated with four attributes, i.e., \textbf{sport appetite}, \textbf{music appetite}, \textbf{work} and \textbf{location}. There are three communities. Left community concentrate on \textbf{music appetite}. People in middle community form friendship due to their \textbf{work} and \textbf{location}. Right community have similar values on \textbf{sport appetite}. Different applications may have different focus attributes and thus require different sets of communities. For example, a marketing manager selling sports goods requires sets of communities with similar \textbf{sport appetite}, and then offers trial products to a few members from each community based on their \textbf{sport appetite} to expect the products to be popular in the communities. A headhunting company may require sets of communities with similar \textbf{work} and \textbf{location} to find the suitable talents.

\begin{figure*}
\centering
\includegraphics{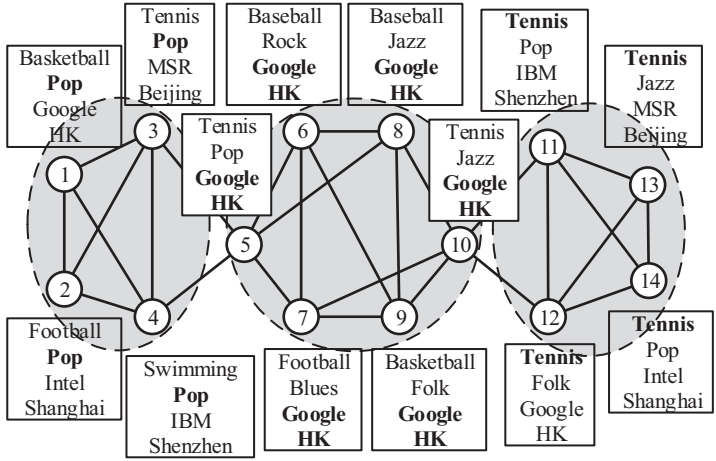}
% where an .eps filename suffix will be assumed under latex,
% and a .pdf suffix will be assumed for pdflatex; or what has been declared
% via \DeclareGraphicsExtensions.
\caption{A example network with three communities.}
\label{F1}
\end{figure*}

In this paper, we detect the set of target communities in the target subspace for any specific application. Determining the target subspace for the specific application is crucial for the problem. FocusCO \cite{Perozzi2014Focused} determines the target subspace by providing a set of exemplar nodes perceived similar by the user. It infers the importnce weights of attributes that capture the user-perceived similarity. However, perceiving a set of exemplar nodes similar by the user is a bit subjective and may not consider the relations between the attribute similarity and the connections. Though the formations of some nodes' connections may be due to their similar attributes, not all similar attributes will lead to cohesive connections which are the basis of the communities. The attributes with large weights that capture the user-perceived similarity may not correlated with the cohesive connections. For example, in Fig. 1, providing $\{3, 5, 11, 14\}$ as exemplar nodes will infer a subspace with \textbf{sport appetite} and \textbf{music appetite} as focus attributes, while such two attributes may not be correlated with the cohesive connections and there is no meaningful communities in the inferred subspace. Thus in order to make the inferred subspace correlated with the cohesive connections, it should be inferred from a set of connected exemplar nodes that had better belong to the same community. FocusCO does not clarify how many exemplar nodes are provided, but definitely more than two. Providing more exemplar nodes accurately is more difficult for an user, especially when limited sample information is available. In order to improve the university of the proposed algorithm, we address the problem in an extreme case where only two sample nodes in any potential target community are provided. When more than two sample nodes are provided, the proposed algorithm is still usefull by selecting two of them as input.

Considering the factors mentioned above, we put forward TSCM, a Target Subspace and Communities Mining method. Two sample nodes in any potential target community of an application are provided by a domain expert. Two sample nodes contain too limited direct information to compute the target subspace accurately. Moreover, they may not have connection relation to make the inferred subspace correlated with the cohesive connections. Thus we design a sample information extension method to extend the two sample nodes to a set of connected exemplar nodes from which we infer the target subspace. Instead of partitioning the whole network, only the set of target communities in the target subspace are mined. The set of target community seeds are extracted by the guidance of the target subspace. Then they are locally expanded to target communities by optimizing a subspace fitness function. Finally, all redundant communities are eliminated. Besides mining the target subspace and communities for specific application, our method has an additional function that it can analyze attribute subspaces and communities around some important node. 

The rest of this paper is organized as follows. Section 2 describes the target subspace communities mining problem. The proposed method TSCM is described in details in section 3. Section 4 discusses related works. Experimental results are presented in section 5. Finally, section 6 concludes the paper.

\section{Problem Formulation}
\label{sec:2}
An attributed network is defined as $\mathcal{G}=(\mathcal{V},\mathcal{E},\mathcal{F})$, where $\mathcal{V}$ is a set of $n$ nodes, $\mathcal{E}$ is the set of $m$ edges and $\mathcal{F}: \mathcal{V}\rightarrow\mathcal{D}_1\times\cdots\times\mathcal{D}_r$ is an attribute function which gives each node an attribute vector $\mathcal{F}(v)$. $Dim=\{1,2,\cdots,r\}$ is the set of all attribute dimensions and $\mathcal{D}_r$ is the value domain of attribute $r$. $\mathcal{F}_t(v)$ denotes the value of attribute $t$ for node $v$. An attribute subspace is represented by a subspace vector $l$. $l_t$ measures the importance weight of the attribute $t$ in the subspace. Since we only consider the relative importance between attributes, the subspace satisfies the normalized condition, i.e., $\sum_{t=1}^r l_t =1,l_t\geq0$.

Many applications do not require all communities of a network. Thus they do not need to partition the whole network. Instead, some specific application usually require a set of target communities whose nodes are similar on some focus attributes. Such property can be captured by a target subspace with large importance weights for the focus attributes. Then the target communities whose nodes are similar on the focus attributes will have large attribute similarity inside under the target subspace. The similarity computed under the target subspace guides the target community mining for the application. Thus the first subproblem is mining the target subspace capturing the essence of an application's requirements. The subproblem is addressed in an extreme case with limited sample information where a domain expert can provide only two sample nodes in any potential target community whose nodes are similar on some focus attributes of a specific application. The subspace inferred from the information of the potential target community containing two sample nodes is adopted as the target subspace capturing the essence of the application's requirements. 

After inferring the target subspace from two sample nodes, the second subproblem is extracting the set of target communities from the network that (1) are structurally dense and well separated from the rest of the network, as well as (2) have large attribute similarity inside under the target subspace. Although the target subspace is inferred from a potential target community, some other communities may also have relatively large similarity inside under the target subspace, as long as their nodes are also similar on the focus attributes. Different communities may strictly match with different subspaces \cite{Mcauley2014Discovering}, but we do not aim to mine the communities strictly matching with the target subspace. Instead, we mine the set of communities as long as they have relatively large similarity inside under the target subspace. In a word, the problem of mining target subspace and communities is defined as follows. Given an attributed network $\mathcal{G}=(\mathcal{V},\mathcal{E},\mathcal{F})$ and two sample nodes $v_{s1}$, $v_{s2}$ from any potential target community of an application, infer the target subspace $l$ of such potential target community and mine the set of target communities $\mathcal{H}$ that (1) are densely intra-connected and sparsely connected with the rest of the network, as well as (2) have large attribute similarity inside under the inferred target subspace. 

The set of target communities are locally extracted rather than obtained by partitioning the whole network. We define a local quality function to evaluate the two requirements of each target community mentioned above. The fitness function \cite{Lancichinetti2009Detecting} is adopted to evaluate the structure cohesiveness of a target community. Let $A=[A_{v,u}]_{v,u=1}^n$ be the adjacency matrix of a network. The fitness of a community $C$ is defined as
\begin{equation}\label{E1}
  fit^C = \frac{invol^C}{vol^C},
\end{equation}
where $invol^C=\sum_{v,u\in C}A_{v,u}$ measures the total internal degrees of the nodes in community $C$, $vol^C=\sum_{u\in C,v\in V}A_{v,u}$ measures the total degrees of the nodes in community $C$. The fitness will get larger value when the community has more edges inside while less edges across the boundary. In order to evaluate the attribute similarity of a target community, we re-weigh the network by setting the attribute similarity under the target subspace as the edge weight and then modify the fitness to subspace fitness accordingly. Assuming that the inferred target subspace is $l$, we adopt the Exponential kernel of attribute vectors as their attribute similarity under $l$:
\begin{equation}\label{E2}
\begin{split}
    s_l(v,u)&=k(||\mathcal{F}(v)-\mathcal{F}(u)||_l)\\
    &=e^{-||\mathcal{F}(v)-\mathcal{F}(u)||_l},
\end{split}
\end{equation}
where $||\mathcal{F}(v)-\mathcal{F}(u)||_l$ is the weighted Euclidean distance under the subspace $l$, i.e., 
\begin{equation}\label{E3}
\begin{split}
    & ||\mathcal{F}(v)-\mathcal{F}(u)||_l \\
    =& \sqrt{(\mathcal{F}(v)-\mathcal{F}(u))^T diag(l)(\mathcal{F}(v)-\mathcal{F}(u))},
\end{split}
\end{equation}
where $diag(l)$ is a diagonal matrix whose main diagonal is $l$. The network is re-weighted as $A^l=[A_{v,u}^l]_{v,u=1}^n$, where
\begin{equation}\label{E4}
  A_{v,u}^l=s_l(v,u)\cdot \mathbb{I}((v,u)\in \mathcal{E}),
\end{equation}
where $\mathbb{I}$ is an indicator function whose value is 1 if the expression inside is true and 0 otherwise. In the re-weighted network, the edge will get larger weight if its two incident nodes are more similar under the target subspace $l$. The subspace fitness is defined on the re-weighted network as
\begin{equation}\label{E5}
  fit_l^C=\frac{\sum_{v,u\in C}A_{v,u}^l}{\sum_{u\in C,v\in \mathcal{V}}A_{v,u}^l},
\end{equation}
The subspace fitness can evaluate the quality of each target community. It gets larger value when the community has more edges inside while less edges across the boundary, as well as has larger attribute similarity inside under the inferred target subspace. 

In order to avoid the mined target communities overlapping heavily, we define a redundancy relationship between two target communities. Given a redundancy parameter $\beta \in [0,1]$, a community $C'$ is redundant with respect to $C$ ($C'\preccurlyeq_{red} C$), if and only if $fit_l^{C'}\leq fit_l^C \wedge \frac{|C'\cap C|}{|C'\cup C|}\geq \beta$. All redundant target communities are eliminated. Smaller redundancy parameter will allow lower overlap between target communities. $\beta=0.5$ is recommended to allow moderate overlap if there is no specific preference.

At least three types of attribute exist in real-world networks, i.e., numerical, binary and categorical attributes. Their value differences are defined uniformly to make them be treated fairly in the computation of the weighted Euclidean distance. The values of each numerical attribute are normalized to the range [0,1]. Then $\mathcal{F}_t(v)-\mathcal{F}_t(u)$ itself denotes the value differences of a numerical attribute $t$. The value difference of a categorical attribute is set as 0 if two values are the same, otherwise 1. For a binary attribute, 1-0 indicate whether a node has such attribute or not. Thus the value difference is set as 0 if two nodes both have such binary attribute, otherwise 1.

\section{Method}
\label{sec:3}
\begin{figure*}
\centering
\includegraphics{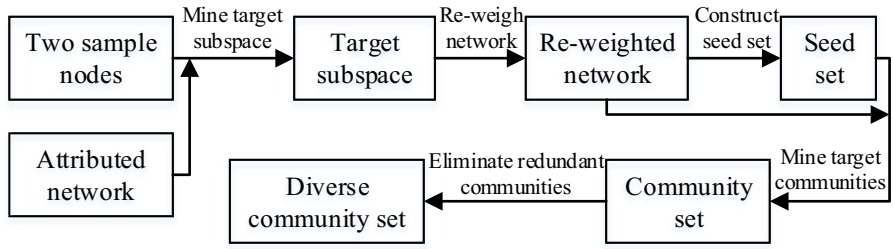}
\caption{Overall framework of TSCM.}
\label{F2}
\end{figure*}
In this section, our method TSCM is described. The overall framework of our method is illustrated in Fig. 2. Given an attributed network and two sample nodes, the method starts from mining the target subspace from two sample nodes. Then the network is re-weighted based on the target subspace and the target community seed set is constructed. Finally, the set of target communities is mined and the redundant ones are eliminated. The overall algorithm will be described first, and then the key procedures will be explained in details.

\subsection{The TSCM Algorithm}
\label{sec:3.1}
The TSCM is given in Algorithm 1. A domain expert steers the method by providing two sample nodes $v_{s1}$, $v_{s2}$ from any potential target community of a specific application. The target subspace is first mined from two sample nodes by $\mathrm{MINE\_TARGET\_SUBSPACE}$ which is explained in details in the next subsection.
\renewcommand{\algorithmicrequire}{\textbf{Input:}}
\renewcommand{\algorithmicensure}{\textbf{Output:}}
%\floatname{algorithm}{Procedure}
\begin{algorithm}
\caption{TSCM}
\label{A1}
\begin{algorithmic}[1]
\REQUIRE attributed network $\mathcal{G}=(\mathcal{V},\mathcal{E},\mathcal{F})$, redundancy parameter $\beta$, and two sample nodes $v_{s1}$, $v_{s2}$.
\ENSURE the target subspace $l$ and the set of diverse target communities $\mathcal{H}$.
 \STATE $\mathcal{H}\leftarrow\emptyset$;
 \STATE $l\leftarrow \mathrm{MINE\_TARGET\_SUBSPACE}(\mathcal{G},v_{s1},v_{s2})$;
 \STATE $[\mathcal{C},A^l]\leftarrow \mathrm{CONSTRUCT\_SEED\_SET}(\mathcal{G},l)$;
 \FOR {each $C\in \mathcal{C}$}
% \IF {$C\nsubseteq visitedNodes$}
 \STATE $C\leftarrow\mathrm{ADJUST\_COMMUNITY}(A^l,C)$;
 \STATE $\mathcal{H}\leftarrow \mathcal{H}\cup\{C\}$; 
% \STATE $visitedNodes\leftarrow visitedNodes\cup C$;
% \ENDIF
 \ENDFOR
 \STATE $\mathcal{H}\leftarrow\mathrm{SELECT\_DIVERSE\_COMMUNITIES}(\mathcal{H},\beta)$;
 \RETURN $l$ and $\mathcal{H}$;
 \end{algorithmic}
\end{algorithm}

Having inferred the target subspace $l$, we mine the set of target communities. We first locate the target communities by identifying the target community seeds that potentially belong to the target communities, and then adjust these seeds to find the target communities. The process of identifying the target community seeds is detailed in Procedure 1. The target community seeds should be densely intra-connected, and have large attribute similarity inside under the target subspace as the target communities do. Thus the network is first re-weighted based on equation (4) where the edge weights are set as the similarity of their end nodes, and the edges with notably large weights are selected to make up a network backbone. All cohesive parts of the network backbone are detected by a community detection algorithm LPA, and they are set as the target community seeds. An edge is deemed to have notably large weight if its weight is larger than the average of the maximum weight and the average weight of the re-weighted network. 
\setcounter{algorithm}{0}
\floatname{algorithm}{Procedure}
\begin{algorithm}
  \caption{$\mathrm{CONSTRUCT\_SEED\_SET}$}
  \label{P1}
  \begin{algorithmic}[1]
    \REQUIRE
    attributed network $\mathcal{G}=(\mathcal{V},\mathcal{E},\mathcal{F})$, target subspace $l$.
    \ENSURE
    target community seed set $\mathcal{C}$, re-weighted network adjacency matrix $A^l$.
    \FOR {each $(v,u) \in \mathcal{E}$}
    \STATE $A^l_{v,u}\leftarrow e^{-\sqrt{(\mathcal{F}(v)-\mathcal{F}(u))^T diag(l)(\mathcal{F}(v)-\mathcal{F}(u))}}$;
    \ENDFOR
    \STATE $A^l_{max}\leftarrow max(A^l)$; $A^l_{avg}\leftarrow mean(A^l)$;
    \FOR {each $(v,u) \in \mathcal{E}$}
    \IF {$A^l_{v,u}\geq \frac{1}{2}(A^l_{max}+A^l_{avg})$}
    \STATE add $(v,u)$ to a network backbone $BB$;
    \ENDIF
    \ENDFOR
    \STATE $\mathcal{C}\leftarrow LPA(BB)$;
    \RETURN $\mathcal{C}$ and $A^l$;
  \end{algorithmic}
\end{algorithm}

Each target community seed is then adjusted to increase its quality function, i.e., subspace fitness defined in equation (5). Inspired by the algorithm in \cite{Lancichinetti2009Detecting}, we adopt a hill-climbing greedy method to locally adjust each community until its subspace fitness can not be increased any more. The process of adjusting community is detailed in Procedure 2. The community seed is set as the initial community. In each iteration, $\mathrm{ADJUST\_COMMUNITY}$ computes the subspace fitness changes of all possible adjustment actions including adding every neighbor to or removing every node from the current community. The action with the largest positive subspace fitness change is selected to modify the community. The iteration continues until no action leads to positive fitness change. The convergence is guaranteed, as each adjustment of the community increases the fitness and the fitness has a maximum value, i.e., 1.
\begin{algorithm}
  \caption{$\mathrm{ADJUST\_COMMUNITY}$}
  \label{P2}
  \begin{algorithmic}[1]
    \REQUIRE re-weighted network adjacency matrix $A^l$, target community seed $C$.
    \ENSURE target community $C$.
    \REPEAT
    \STATE $\Delta f_{best}\leftarrow0$;
    \STATE $Actions\leftarrow\{\mathrm{REMOVE}(v)| v\in C\}\cup\{\mathrm{ADD}(v)|v\in \mathcal{V}\setminus C \wedge \exists u\in C:(v,u)\in \mathcal{E}\}$;
    \FOR {each $a\in Actions$}
    \STATE $\Delta f\leftarrow \mathrm{COMPUTE\_\Delta\_FITNESS}(A^l,a,C)$;
    \IF {$\Delta f>\Delta f_{best}$}
    \STATE $\Delta f_{best}\leftarrow \Delta f$; $bestAction\leftarrow a$;
    \ENDIF
    \ENDFOR
    \IF {$\Delta f_{best}>0$}
    \STATE $C\leftarrow \mathrm{MODIFY}(C,bestAction)$;
    \ENDIF
    \UNTIL {$\Delta f_{best}=0$}
    \RETURN $C$;
  \end{algorithmic}
\end{algorithm}

After handling all seeds, we eliminate all redundant target communities by $\mathrm{SELECT\_DIVERSE\_COMMUNITIES}$. All detected communities are checked one by one according to descending order of their subspace fitness. If the checked one is not redundant to any one in current diverse communities set, it is added to the set. Finally the target subspace and diverse target communities set are returned as the output.

\subsection{Mining Target Subspace}
\label{sec:3.2}
The target subspace is inferred from two sample nodes in any potential target community whose nodes are similar on some focus attributes of a specific application. Two sample nodes can only form one pair of nodes, and it is hard to determine the focus attributes from them directly. We expect to expand the two sample nodes to a set of exemplar nodes which are in the potential target community and similar to each other on those focus attributes. The neighbors of two sample nodes are taken as candidates for sample nodes expansion, because (1) some neighbors must also be in the target community that contains the sample nodes and similar to sample nodes on the focus attributes, as well as (2) the connection relations between the neighbors and the sample nodes make the inferred subspace more likely to be correlated with cohesive connections. However, not all neighbors of the sample nodes come from the target community containing the sample nodes, as sample nodes may connect with nodes in some other communities. According to structure cohesiveness property of the community, neighbors belonging to the same community should have relatively dense connections. We define the neighborhood network of a node as follows. 
\begin{definition}[Neighborhood Network]
Given a node $v$ whose neighbors set is denoted as $NB(v)=\{w|(v,w)\in \mathcal{E}\}$, the neighborhood network of node $v$ is defined as $NN(v)=(NB(v),NE(v))$, where the edge set $NE(v)=\{(u,w)|u\in NB(v)\wedge w\in NB(v)\wedge (u,w)\in \mathcal{E}\}$.
\end{definition}
Then the neighbors belonging to the same community should be cohesive in the corresponding neighborhood network. The cohesive parts of a neighborhood network can also be detected by a community detection method LPA. The cohesive parts of the neighborhood network of a node is called the neighborhood communities of such node. It is reasonable to conclude that at least one neighborhood community of each sample node comes from the target community containing the sample nodes. We need to decide which neighborhood communities come from that target community and then use them to expand the sample nodes. 

Intuitively, nodes in the different neighborhood communities from the same target community should be similar to each other on the same set of focus attributes, respectively. Each neighborhood community plus its corresponding sample node is set as a set of exemplar nodes. For each set of exemplar nodes, an attribute subspace that makes them similar to each other is computed. The subspace computed from the set of exemplar nodes composed of the neighborhood community $nc$ is deemed as the subspace of such neighborhood community $nc$. Then the subspaces of those neighborhood communities from the same target community should have large similarity, as they have similar attributes with large importance weights. The similarity of two subspaces is defined based on cosine similarity, i.e., 
\begin{equation}\label{E6}
  SS=\frac{l_1\cdot l_2}{|l_1||l_2|};
\end{equation}
where $l_1$ and $l_2$ are two subspaces. Assuming that $\mathcal{L}(v)$ is the set of subspaces of the neighborhood communities of a node $v$, the similarity between each subspace in $\mathcal{L}(v_{s1})$ of sample node $v_{s1}$ and that in $\mathcal{L}(v_{s2})$ of sample node $v_{s2}$ is computed. Then the subspace of the neighborhood community from the target community in $\mathcal{L}(v_{s1})$ and that in $\mathcal{L}(v_{s2})$ should have the largest similarity. In this way, we locate the neighborhood community that comes from the target community for each sample node. Such two neighborhood communities and two provided sample nodes make up the final set of exemplar nodes from which we infer the target subspace. 

Take the toy network in Fig. 1 as an example. Assuming that the middle community is a potential target community whose subspace has focus attributes \textbf{work} and \textbf{location}, and nodes 5 and 10 are two provided sample nodes, it is hard to determine the focus attributes from 5 and 10 directly. some neighbors of two sample nodes are also in the target community such as 6 and 8, etc., but some are in other communities. Node 5 has two neighborhood communities $\{3,4\}$ and $\{6,7,8\}$. Node 10 has two neighborhood communities $\{11,12\}$ and $\{7,8,9\}$. Neighborhood communities $\{6,7,8\}$ and $\{7,8,9\}$ are from the target community. Nodes in both neighborhood communities are similar to each other on attributes \textbf{work} and \textbf{location}. After computing the similarities of subspaces between two sample nodes, the subspaces of $\{6,7,8\}$ and $\{7,8,9\}$ will have the largest similarity. In this way, $\{6,7,8\}$ and $\{7,8,9\}$ are selected to form the final set of exemplar nodes. The final set of exemplar nodes is $\{5,6,7,8,9,10\}$.

The process of mining the target subspace is detailed in Procedure 3. $\mathrm{DETECT\_NEI\_COMMUNITY}$ detects all neighborhood communities in neighborhood network of a node $v$. For each neighborhood community, computing its subspace. After computing the subspaces of neighborhood communities of two sample nodes, we compute their similarities based on equation (6). Two subspaces with largest similarity is selected. Their corresponding sets of exemplar nodes are selected to form the final set of exemplar nodes. The target subspace is computed from the final set of exemplar nodes.
\begin{algorithm}
  \caption{$\mathrm{MINE\_TARGET\_SUBSPACE}$}
  \label{P3}
  \begin{algorithmic}[1]
    \REQUIRE
    attributed network $\mathcal{G}=(\mathcal{V},\mathcal{E},\mathcal{F})$, two sample nodes $v_{s1}$, $v_{s2}$.
    \ENSURE
    target subspace $l$.
    \FOR {each $v \in \{v_{s1},v_{s2}\}$}
    \STATE $\mathcal{L}(v)\leftarrow\emptyset$;
    \STATE $NC\leftarrow \mathrm{DETECT\_NEI\_COMMUNITY}(\mathcal{G},v)$;
    \FOR {each $nc\in NC$}
    \STATE $\mathcal{T}\leftarrow nc\cup \{v\}$;
    \STATE $l\leftarrow \mathrm{COMPUTE\_SUBSPACE}(\mathcal{G},\mathcal{T})$;
    \STATE $\mathcal{L}(v)\leftarrow\mathcal{L}(v)\cup \{l\}$;
    \ENDFOR
    \ENDFOR
    \FOR {each $l_1 \in \mathcal{L}(v_{s1})$}
    \FOR {each $l_2 \in \mathcal{L}(v_{s2})$}
    \STATE $SS(l_1,l_2)\leftarrow\frac{l_1\cdot l_2}{|l_1||l_2|}$;
    \ENDFOR
    \ENDFOR
    \STATE $(l_1^{*},l_2^{*})\leftarrow\max_{(l_1,l_2)}SS(l_1,l_2)$;
    \STATE Select $\mathcal{T}_1^{*}$ and $\mathcal{T}_2^{*}$ corresponding to $l_1^{*}$ and $l_2^{*}$, respectively;
    \STATE $\mathcal{T}\leftarrow\mathcal{T}_1^{*}\cup\mathcal{T}_2^{*}$;
    \STATE $l\leftarrow \mathrm{COMPUTE\_SUBSPACE}(\mathcal{G},\mathcal{T})$;
    \RETURN $l$;
  \end{algorithmic}
\end{algorithm}

We now discuss how to compute the subspace from a set of exemplar nodes. The computed subspace should make the exemplar nodes similar to each other. In other words, it should make the exemplar nodes have small distance to each other. The distance is measured by equation (3). FocusCO infers the subspace from a set of exemplar nodes by optimizing a distance metric learning objective function. However, optimizing its objective function is time consuming and the optimal solution is hard to get, due to the high dimensionality and the positive semidefinite constraints of the optimization problem. Thus we aim to design a direct calculation method rather than optimization method to infer the subspace. Assuming that $\mathcal{T}$ is a set of exemplar nodes, $P_S$ is the set of all pairs of exemplar nodes, and $P_R$ is the set of node pairs randomly sampled from the whole network excluding the set of exemplar nodes, i.e., $\mathcal{V}\setminus\mathcal{T}$. Pairs in $P_S$ have small distance under the subspace of $\mathcal{T}$. In order to ensure that $P_R$ contains enough dissimilar node pairs that have large distance under the subspace of $\mathcal{T}$, the size of $P_R$ is set as $r|P_S|$, where $r$ is the number of the attributes. For a node pair set $P$, the average square difference of attribute $t$ between node pairs in it is denoted as
\begin{equation*}
h_t(P)=\frac{1}{|P|}\sum_{(v,u)\in P}(\mathcal{F}_t(v)-\mathcal{F}_t(u))^2;
\end{equation*}
The subspace $l$ of the exemplar nodes set $\mathcal{T}$ should make the average distance between pairs in $P_S$, i.e. $\sum_{t=1}^r l_t h_t(P_S)$, relatively small, while make that in $P_R$, i.e. $\sum_{t=1}^r l_t h_t(P_R)$, relatively large. If $h_t(P_S)$ is no smaller than $h_t(P_R)$, the importance weight $l_t$ is set as 0, as attribute $t$ has no contribution to make the average distance between pairs in $P_S$ relatively small while make that in $P_R$ relatively large. Otherwise, if $h_t(P_S)$ is smaller than $h_t(P_R)$, the importance weight $l_t$ is set proportinoal to $\frac{h_t(P_R)}{h_t(P_S)+1/|P_S|}$, as attribute $t$ is more important for subspace when pairs in $P_S$ are more similar than that in $P_R$ on such attribute. Considering the normalized condition of the subspace, the subspace $l$ of $\mathcal{T}$ can be directly calculated as
\begin{equation*}
  l_t^\prime=\left\{
  \begin{array}{rl}
  \!\frac{h_t(P_R)}{h_t(P_S)+1/|P_S|}, &\ if\quad h_t(P_S) < h_t(P_R)\\
  \!0, &\ Otherwise
  \end{array}
  , t=1,2,\cdots,r;
  \right.
\end{equation*}
\begin{equation}\label{E7}
  l_t=\frac{l_t^\prime}{\sum_{t=1}^r l_t^\prime},\; t=1,2,\cdots,r;
\end{equation}

The process of computing the subspace of a set of exemplar nodes is detailed in Procedure 4. $\mathrm{GET\_ALL\_PAIRS}$ returns all pairs of exemplar nodes in $\mathcal{T}$. $\mathrm{SAMPLE\_RANDOM\_PAIRS}$ randomly samples $r|P_S|$ node pairs from $\mathcal{V}\setminus\mathcal{T}$. For each attribute $t$, $h_t(P_S)$ and $h_t(P_R)$ are first computed, and then $l_t^\prime$ is computed based on the relations between $h_t(P_S)$ and $h_t(P_R)$. Finally, the subspace $l$ is computed by normalizing $l^\prime$.
\begin{algorithm}
  \caption{$\mathrm{COMPUTE\_SUBSPACE}$}
  \label{P4}
  \begin{algorithmic}[1]
    \REQUIRE
    attributed network $\mathcal{G}=(\mathcal{V},\mathcal{E},\mathcal{F})$, the set of exemplar nodes $\mathcal{T}$.
    \ENSURE
    subspace $l$.
    \STATE $P_S\leftarrow\mathrm{GET\_ALL\_PAIRS}(\mathcal{T})$;
    \STATE $P_R\leftarrow\mathrm{SAMPLE\_RANDOM\_PAIRS}(\mathcal{V},\mathcal{T},r|P_S|)$;
    \FOR {$t=1$ to $r$}
    \STATE $h_t(P_S)=\frac{1}{|P_S|}\sum_{(v,u)\in P_S}(\mathcal{F}_t(v)-\mathcal{F}_t(u))^2$;
    \STATE $h_t(P_R)=\frac{1}{|P_R|}\sum_{(v,u)\in P_R}(\mathcal{F}_t(v)-\mathcal{F}_t(u))^2$;
    \IF {$h_t(P_S) < h_t(P_R)$}
    \STATE $l_t^\prime=\frac{h_t(P_R)}{h_t(P_S)+1/|P_S|}$;
    \ELSE
    \STATE $l_t^\prime=0$;
    \ENDIF
    \ENDFOR
    \STATE $l=\frac{l^\prime}{\sum_{t=1}^r l_t^\prime}$;
    \RETURN $l$;
  \end{algorithmic}
\end{algorithm} 

\subsection{Computational Complexity}
\label{sec:4.3}
Given an attributed network with $n$ nodes, $m$ edges and $r$ attributes, the time complexity of TSCM is analyzed as follows.

We first analyze the time complexity of $\mathrm{MINE\_TARGET\_SUBSPACE}$. $\mathrm{DETECT\_NEI\_COMMUNITY}$ requires about $O(\overline{d})$ time to detect neighborhood communities by a linear time community detection method, where $\overline{d}$ denotes the average degree of node. In $\mathrm{COMPUTE\_SUBSPACE}$, $\mathrm{GET\_ALL\_PAIRS}$ takes $O(|\mathcal{T}|^2)$ time, $\mathrm{SAMPLE\_RANDOM\_PAIRS}$ takes $O(r|\mathcal{T}|^2)$ time, computing $h_t(P_S)$, $h_t(P_R)$ and $l_t^\prime$ take $O(r|\mathcal{T}|^2)$ time for each attribute $t$, and computing the subspace $l$ takes $O(r)$ time. Thus $\mathrm{COMPUTE\_SUBSPACE}$ totally requires $O(r^2|\mathcal{T}|^2)$ time. Assuming that the average neighborhood community size is $\overline{c}$, and the average number of neighborhood communities of a node is $\overline{a}$, then $\overline{d}\approx\overline{a}\overline{c}$, $|\mathcal{T}|=O(\overline{c})$. It takes $O(\overline{a}^2 r)$ time to compute all subspace similarities, and it takes $O(\overline{a}^2)$ time to select two subspaces with largest similarity. Based on the operational rules of the symbol $O$, the total time complexity of $\mathrm{MINE\_TARGET\_SUBSPACE}$ is $O(\overline{a}r^2\overline{c}^2+\overline{a}^2 r)$.

In $\mathrm{CONSTRUCT\_SEED\_SET}$, it first takes $O(mr)$ time to re-weigh the network, computing the maximum weight and the average weight takes $O(m)$ time, the network backbone is extracted with time $O(m)$ and the community seeds are detected by a linear time community detection method with time $O(m)$. Thus $\mathrm{CONSTRUCT\_SEED\_SET}$ requires $O(mr)$ time. $\mathrm{ADJUST\_COMMUNITY}$ takes $O(\overline{|C|} n \overline{d})$ time where $\overline{|C|}$ is the average target community size, because the computation of a subspace fitness change of an action averagely requires $O(\overline{d})$, at most $n$ possible actions are taken to adjust one node of the community, and about $\overline{|C|}$ nodes need to be adjusted to form the final community. In $\mathrm{SELECT\_DIVERSE\_COMMUNITIES}$, assuming that the number of original target communities and the number of diverse target communities are $k$ and $h$ respectively, then all original target communities are sorted with time $O(k\log k)$, and each original target community is checked with all diverse target communities with time at most $O(h\overline{|C|})$. Thus $\mathrm{SELECT\_DIVERSE\_COMMUNITIES}$ takes $O(k\log k+k h\overline{|C|})$ which is further simplified to $O(k^2\overline{|C|})$. To sum up, the total time complexity is $O(\overline{a}r^2\overline{c}^2+\overline{a}^2 r+mr+k\overline{|C|} n \overline{d}+k^2\overline{|C|})$. $\overline{c}$, $\overline{a}$, $\overline{d}$ usually do not increase as the network size increases and are far smaller than the network size, so they are regarded as constants in time complexity. Based on the operational rules of the symbol $O$, the total time complexity is simplified to $O(r^2+m r+k\overline{|C|} n+k^2\overline{|C|})$. It is worth noting that $k\overline{|C|}$ is smaller than $n$, since our algorithm only extracts the set of target communities in the target subspace rather than partitioning the whole network. Moreover, $k\overline{|C|}$ may not increase as $n$ gets larger, as $\overline{|C|}$ is usually independent of $n$, and $k$ mainly depends on the specific application. 

\subsection{Discussions}
\label{sec:4.4}
We discuss a couple of variants of our problem setting and how to adapt the proposed algorithm to deal with these variants. 

When more than two sample nodes are provided, any two of them can be selected as the input of the TSCM. Moreover, the algorithm can be slightly adapted to take all provided nodes as the input. The main adaptation is conducted on Procedure 3 $\mathrm{MINE\_TARGET\_SUBSPACE}$. For each provided sample node, the subspaces of its all neighborhood communities are computed. Then we randomly select two sample nodes as two prototype nodes and compute similarities of their subspaces. Two subspaces with largest similarity are selected as two prototype subspaces. For each remaining sample nodes except two prototype nodes, we compute the similarities between its subspaces and two prototype subspaces, and then select its subspace with largest total similarity with two prototype subspaces as its feature subspace. All neighborhood communities corresponding to the feature subspaces and the prototype subspaces are likely to be in the potential target community, so they are merged to construct the final set of exemplar nodes from which the target subspace is computed. When more sample nodes are provided, the exemplar node set will be larger and the computed target subspace will be more accurate. 

Besides mining the target subspace and the set of target communities, the proposed TSCM method can also be adapted to analyze the subspaces and communities around an important node by setting such node as the single sample node. The subspaces of all neighborhood communities of such sample node are computed. For each computed subspace, the network is re-weighted based on it by equation (4), its corresponding neighborhood community is set as its community seed, and then its community is extracted by expanding its seed on the re-weighted network to optimize the subspace fitness. In this way, the communities and their subspaces around the sample node are mined.

\section{Related Works}
\label{sec:4}
Most of attribute community detection methods take unsupervised clustering techniques. In the early stage, they determine the attribute homogeneity of each community in attribute full space which treats all available attributes as equally important. The SA-Cluster \cite{Zhou2009Graph} and its extended version Inc-Cluster \cite{Zhou2010Clustering} proposed by Zhou et al. take a K-Medoids method to cluster the augmented attributed network where all attributes are regarded as attribute vertices. CESNA method \cite{Yang2013Community} statistically models the network structure and all node attributes, and gets the communities by optimizing a combined likelihood. Similarly, BAGC method \cite{Xu2012A} adopts a Bayesian model to model both structure and all attributes and obtains communities by inferring parameters of the model. PICS method \cite{Akoglu2012Pics} detects communities by optimizing the total encoding cost which combines model description cost and data description cost of both adjacency matrix and attribute matrix. With the increasing dimensionality of attribute space, the discrimination power of attribute distance or similarity metrics may decrease in full space \cite{Gunnemann2014Gamer}. Thus attribute subspace methods extracting communities with homogeneous attribute values in attribute subspaces are introduced. CoPaM method \cite{Moser2009Mining} mines the set of maximal cohesive patterns defined as a dense and connected subgraph that has homogeneous values in a large enough attribute subspace. SCPM method \cite{Silva2012Mining} extracts set of structure correlation pattern defined as a dense subgraph induced by a particular attribute subset. SSCG \cite{Gunnemann2013Spectral} is a unsupervised spectral subspace clustering method which detects for each cluster an individual subset of relevant attributes and adopts spectral clustering to learn the community structure. GAMer method \cite{Gunnemann2014Gamer} combines the paradigms of dense subgraph mining and subspace clustering to mine the maximal twofold clusters by some pruning strategies. SCMAG method \cite{Huang2015Dense} identifies cells with dense connectivity in the subspaces and uses a cell-based subspace clustering approach to detect the cell-based communities. Unsupervised methods mentioned above globally detect communities but not aim at mining the target subspace and set of target communities for specific application.

Different from unsupervised methods, DCM \cite{Pool2014Description} and FocusCO \cite{Perozzi2014Focused} are two semi-supervised clustering methods whose detection results can be steered by a domain expert. DCM \cite{Pool2014Description} aims to find a set of cohesive communities with concise descriptions from a set of candidate communities provided by a domain expert. The community descriptions defined in DCM are queries consisting of disjunctions of conjunctions over basic conditions on the attribute vectors. They are different from general attribute subspaces considered in this paper. Moreover, a domain expert controls the detection results by providing candidate communities in DCM rather than sample nodes, and one initial candidate community can only lead to one final community. FocusCO \cite{Perozzi2014Focused} allows user to steer the communities by providing a small set of exemplar nodes that are deemed to be similar to one another as well as similar to the type of nodes the communities of his interest should contain. They do not clarify how many exemplar nodes are required, but definitely require more than two exemplar nodes. In this paper, we address the problem in an extreme case where only two sample nodes in any potential target community are provided. 

\section{Experimental Results}
\label{sec:5}
In this section, we thoroughly evaluate the effectiveness and efficiency of TSCM on synthetic networks and show its application values on real-world networks.

\subsection{Evaluation on Synthetic Networks}
\label{sec:5.1}
The experiments settings including synthetic attributed network generation process, the comparison methods and the quality indicators are first described. Then the effectiveness and efficiency comparison results are analyzed.

\subsubsection{Experiments Settings}
Synthetic attributed networks with ground truth communities are generated based on the LFR benchmarks \cite{Lancichinetti2009Benchmarks}. The LFR benchmarks have similar features to real-world networks. Their degree and community size distributions are governed by power laws with exponents $\tau_1$ and $\tau_2$, respectively. The benchmarks are controlled by several other parameters, i.e., node number $n$, average node degree $d_{avg}$, maximum node degree $d_{max}$, minimum community size $c_{min}$, maximum community size $c_{max}$ and mixing parameter $\mu$. Mixing parameter controls the fuzzy degree of the network. The larger the value of $\mu$ is, the fuzzier the benchmark becomes. In order to obtain attributed benchmarks, the attribute vectors are further attached to all nodes. Three types of attributed benchmarks are generated by attaching three types of attribute vectors, i.e., numerical, binary and categorical, respectively. We select the most suitable type of benchmark for each method to evaluate its performance. The attached attribute vectors are controlled by four parameters, i.e., total attribute number $r$, attribute subspace size $t$, target community number $b$ and similarity probability $p$. The subspace size measures the number of focus attributes with large importance weights in each subspace. A random subspace is generated as the target subspace, and $b$ communities are selected as the target communities which are assigned such target subspace. The remaining communities are assigned random subspaces. The importance weights of any subspace are difficult to control in synthetic networks. We consider a simplified case where in each subspace $l$, all focus attributes are assumed to have the same importance weight and the importance weights of other attributes are set as 0, i.e., 
\begin{equation*}
  l_i=\left\{
  \begin{array}{rcl}
  \frac{1}{t}, && \mathrm{if\ } i \mathrm{\ is\ a\ feature\ attribute} \\
  0, &&\mathrm{otherwise}
  \end{array}
  ,
  \right.
\end{equation*}
In this case, all nodes in a community have similar values on each of its focus attribute with probability $p$, while have random values on other attributes. The larger the value of $p$ is, the more homogeneous the community will be on its focus attributes. The default parameters are set as follows, $\tau_1=2$, $\tau_2=1$, $n=5000$, $d_{avg}=30$, $d_{max}=100$, $c_{min}=40$, $c_{max}=2c_{min}=80$, $\mu=0.2$, $r=20$, $t=6$, $b=5$, $p=0.9$. Six sets of benchmarks are generated by separately varying $n$, $c_{min}$, $\mu$, $r$, $t$, and $p$ respectively, while fixing the other parameters.

A set of related community detection methods is selected to compare with TSCM. Louvain \cite{Vincent2008Fast} is a fast method only using network structure. BAGC \cite{Xu2012A} and PICS \cite{Akoglu2012Pics} are two attribute full space methods. They are designed for categorical attribute networks and binary attribute network, respectively. BAGC requires possible maximum community number as input. We run it with 1 to 5 times of the real community number as input respectively, and report the best result. GAMer \cite{Gunnemann2014Gamer} is an attribute subspace method and designed for numerical attribute networks. FocusCO \cite{Perozzi2014Focused} is the only method addressing the similar problem to ours. It requires a user to provide several similar nodes as exemplar nodes. The size of its exemplar set is set as 8 which is much larger than the number of provided sample nodes in TSCM. FocusCO and TSCM can run on all three types of benchmarks and their results on three types of benchmark are marked with suffixes `-num', `-bin', and `-cate', respectively. The results of FocusCO average over 20 runs with randomly provided exemplar nodes for each run, and similar for TSCM. All other parameters of the methods are set as default described in their papers.

Any two nodes in any target community are set as the sample nodes for TSCM. For FocusCO, it requires its exemplar nodes to be similar under the target subspace. Since nodes in any target community have high potential to be similar to each other under the target subspace, the exemplar nodes of FocusCO are selected from a target community. It is obviously that providing two sample nodes is much easier than providing several exemplar nodes from a target community. Similar to TSCM, FocusCO has an attribute subspace inference procedure which infers the subspace where the exemplar nodes are similar to each other. Since the exemplar nodes of FoucsCO are selected from a target community, its inferred subspace should also be as similar to the target subspace as possible. We first compare the qualities of subspaces mined by TSCM and FocusCO. The subspace mining quality is measured by the subspace similarity $SS$ between the mined subspace and the target one. $SS$ is defined in equation (6). The larger the subspace similarity is, the more similar to the target subspace the mined subspace is. The target community detection quality of each method is measured via a quality indicator $Q$ defined based on F1 score. The goal of our problem is to extract all target communities in the target subspace. Let $\mathcal{P}=\{P_i\}$ denote the set of ground truth target communities. Let $\mathcal{R}=\{R_j\}$ denote the set of communities detected by any method. For each method, the F1 score between each community $P_i$ in $\mathcal{P}$ and each community $R_j$ in $\mathcal{R}$, $F1(P_i,R_j)$, is computed. Since our goal is mining all communities in $\mathcal{P}$ accurately, the quality of mining each community $P_i$ in $\mathcal{P}$ is measured by the maximum F1 score between $P_i$ and all detected communities in $\mathcal{R}$, i.e., $QI(P_i) = \max_{R_j\in \mathcal{R}}F1(P_i,R_j)$. The quality indicator $Q$ of a method is defined as the average quality of mining all communities in $\mathcal{P}$, i.e., $Q=\sum_{P_i\in \mathcal{P}}QI(P_i)/|\mathcal{P}|$. The larger the $Q$ is, the better the method can extract all target communities.

\subsubsection{Results Analysis}
\begin{figure*}
\centering
\includegraphics{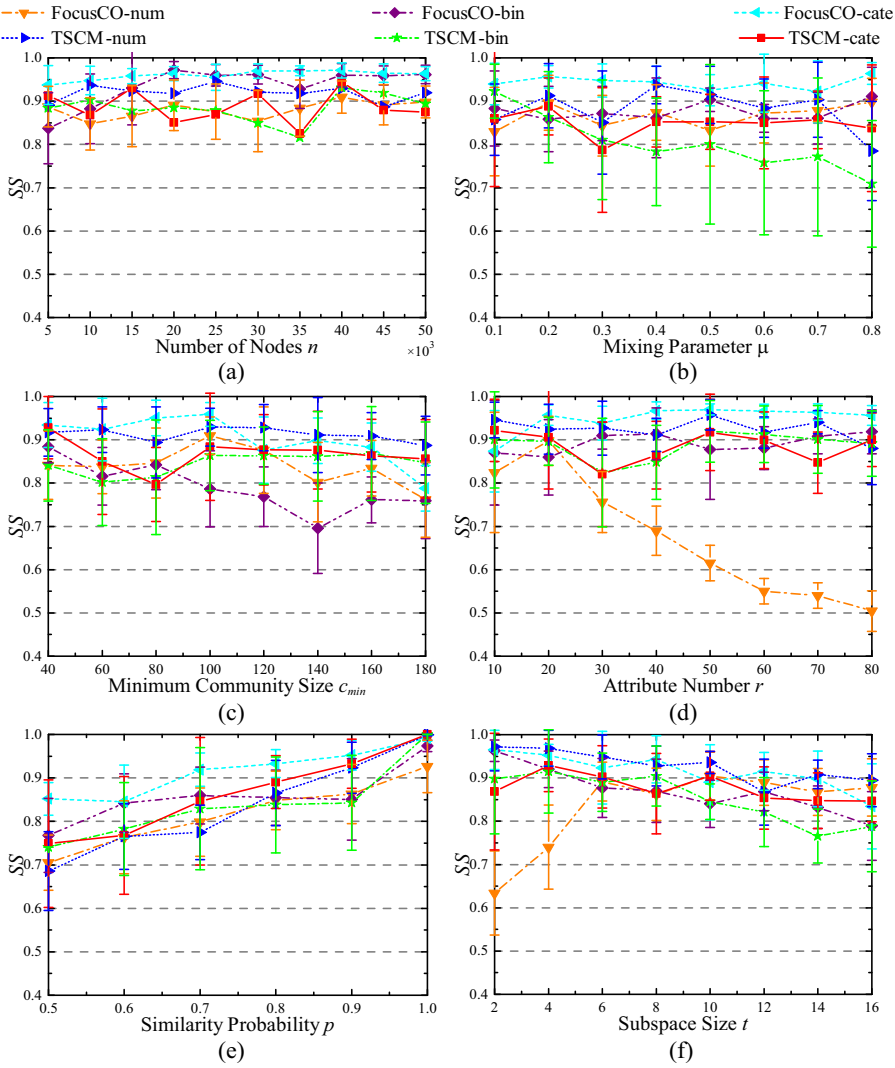}
% where an .eps filename suffix will be assumed under latex,
% and a .pdf suffix will be assumed for pdflatex; or what has been declared
% via \DeclareGraphicsExtensions.
\caption{$SS$ vs. (a) $n$, (b) $\mu$, (c) $c_{min}$, (d) $r$, (e) $p$, (f) $t$. Bars depict standard deviations.}
\label{F3}
\end{figure*}

Fig. 3 shows the subspace mining quality of TSCM and FocusCO on six sets of benchmarks. Both of them keep their subspace similarity around 0.9 in most cases except a few cases described below. In Fig. 3(b), TSCM-bin slightly decreases the subspace similarity as the mixing parameter increases. In Fig. 3(c), FocusCO-bin slightly decreases the subspace similarity as the minimum community size increases. In Fig. 3(d), FocusCO-num decreases the subspace similarity a lot as attribute number gets larger. In Fig. 3(e), both methods increase their similarities as the similarity probability increases. This is because the exemplar nodes of TSCM and FocusCO become more similar to each other under the subspace as the similarity probability increases. In Fig. 3(f), FocusCO-num has small subspace similarity when the subspace size is very small. The subspace mining quality depends on many factors, such as network structure, attribute types, initial provided sample nodes or exemplar nodes and the subspace mining procedures, etc.. Though the number of exemplar nodes in FocusCO is much larger than that of sample nodes in TSCM, TSCM has comparable subspace mining performance with FocusCO. This shows that the sample information extension technique and the subspace mining procedure designed in TSCM is effective.

\begin{figure*}
\centering
\includegraphics{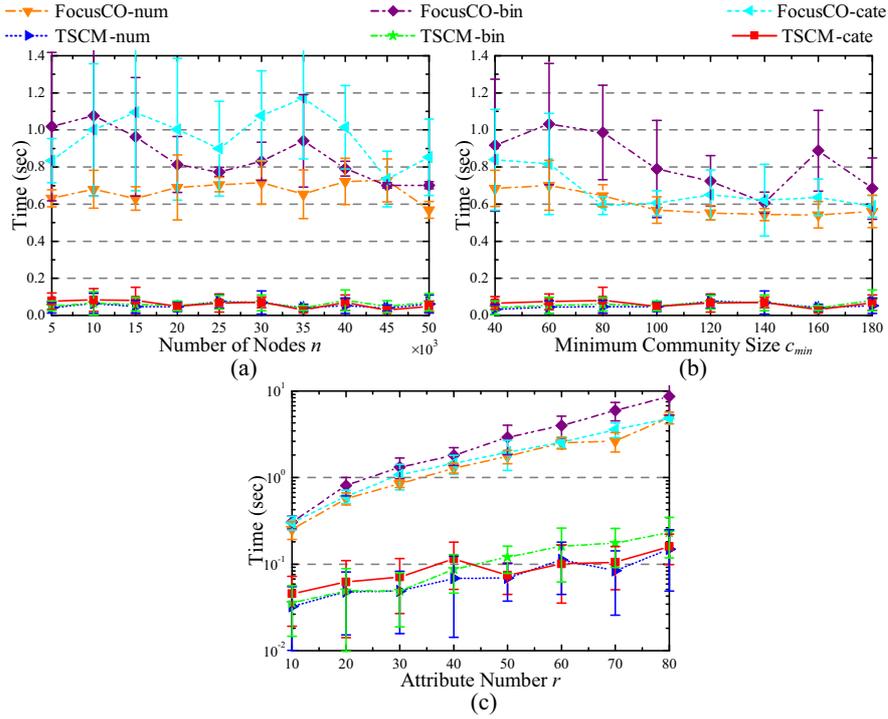}
\caption{Running time of mining target subspace vs. (a) $n$, (b) $c_{min}$, (c) $r$. Bars depict standard deviations.}
\label{F4}
\end{figure*}

The efficiency comparison of subspace mining procedures of TSCM and FocusCO is shown in Fig. 4. The procedure of TSCM is one order of magnitude faster than that of FocusCO in all cases. FocusCO conducts one time of distance metric optimization to compute the subspace, while TSCM requires multiple subspace computations to obtain the final subspace. Thus our subspace computation method based on direct calculation is much faster than that of FocusCO based on distance metric optimization. 

\begin{figure*}
\centering
\includegraphics{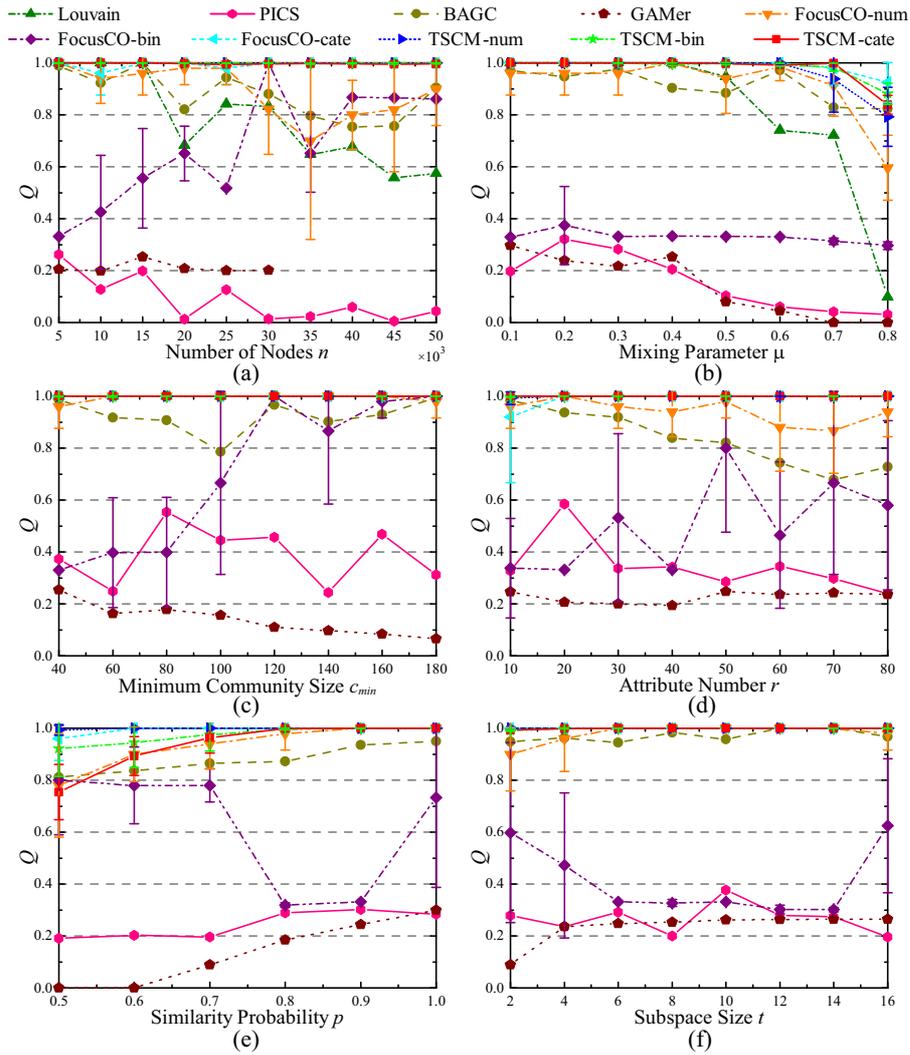}
\caption{$Q$ vs. (a) $n$, (b) $\mu$, (c) $c_{min}$, (d) $r$, (e) $p$, (f) $t$. Bars depict standard deviations. GAMer can't obtain results on networks larger than 30000, due to an out of memory problem.}
\label{F5}
\end{figure*}

Fig. 5 shows the target community mining quality comparisons on six sets of benchmarks. Fig. 5(a) shows the quality changes along with the network size. The quality of Louvain decreases, because the modularity maximization has the resolution limitation problem in large networks. The quality of PICS decreases because it tends to detect relatively large communities, while the larger network makes the community smaller. FocusCO-bin has relatively poor quality especially when the network is small. This is because the constructed community seeds of FocusCO do not have cohesive connection property and easily include nodes from different communities in binary attribute networks. TSCM can almost detect all target communities perfectly in all cases. Fig. 5(b) shows the quality changes for an increasing mixing parameter. When the mixing parameter gets larger, the quality of Louvain decreases a lot, while that of others decreases moderately. All methods except Louvain make use of attribute information when detecting communities. This demonstrates that the attribute information helps community detection in fuzzy networks. Fig. 5(c) shows the quality changes versus community size. GAMer decreases its quality, as it tends to extract small communities. The quality of FocusCO-bin increases, because its community seeds are harder to include nodes from different communities when communities become larger. Fig. 5(d) shows the quality changes along with the attribute number. The quality of BAGC decreases, as it is a full space method, and more attributes make the subspace farther from the full space. FocusCO-num decreases its quality, because its mined subspace becomes less similar to the target one as shown in Fig. 3(d). In Fig. 5(e), most methods increase their quality as the similarity probability increases. Finally, Fig. 5(f) shows the quality changes versus subspace size. BAGC increases its quality slightly, as larger subspace is more close to the full space. In general, on all six sets of benchmarks, TSCM has almost perfect quality nearly all the time except on the benchmarks with low similarity probabilities. FocusCO-bin has poor performances in most cases, as its community seeds easily include nodes from different communities. FocusCO-num is always worse than TSCM-num. Only FocusCO-cate is comparable with TSCM-cate. PICS and GAMer always have very poor quality. The major reason for poor performance of other methods is that they are unsupervised and not specially developed for target subspace and communities mining problem.

\begin{figure*}
\centering
\includegraphics{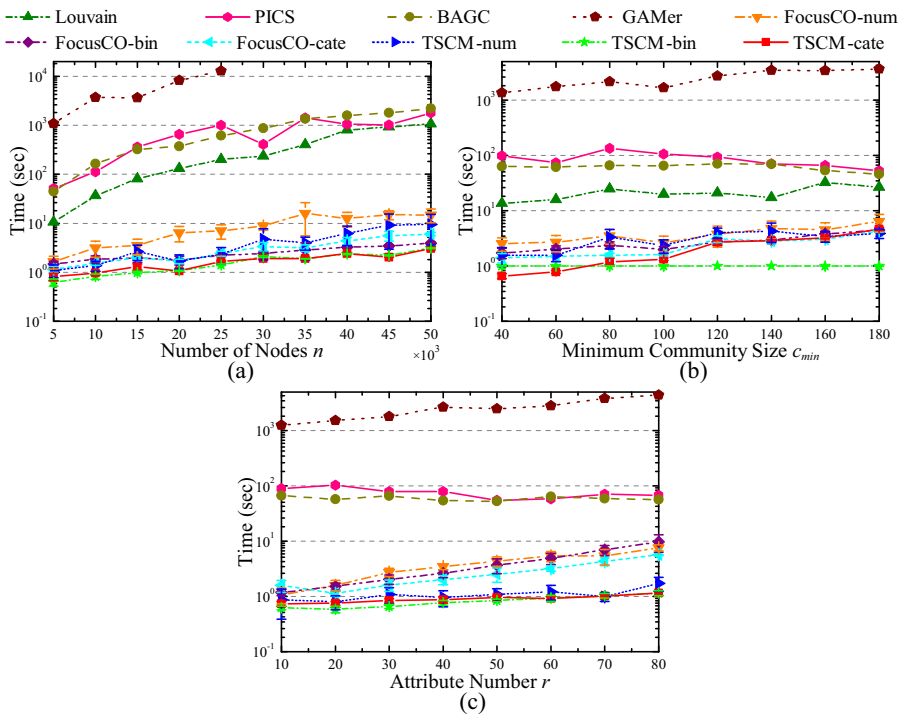}
\caption{Running time vs. (a) $n$, (b) $c_{min}$, (c) $r$. Bars depict standard deviations.}
\label{F6}
\end{figure*}

Finally, Fig. 6 shows the methods' running time. In Fig. 6(a), all methods require more time when the networks get larger. In Fig. 6(b), GAMer, FocusCO and TSCM require more time when the communities get larger, while community size has little influence on the running time of PICS, BAGC and Louvain. In Fig. 6(c), GAMer, FocusCO and TSCM cost more time when the attribute number gets larger, while attribute number has little influence on the running time of PICS, BAGC and Louvain. In general, GAMer costs the most time and is followed by PICS and BAGC. TSCM and FocusCO are several orders of magnitude faster than other methods. This is because they only extract required communities rather than partition the whole networks.

\subsection{Applications on Real-world Networks}
Since there is no ground truth about the target subspace and the set of target communities given for real-world networks and most of compared methods are not designed for target subspace and communities mining problem, it is inherently hard to quantitatively analyze our method on real-world networks. Our following case studies on real-world networks mainly illustrate application values of TSCM. arXiv \footnote{http://www.cs.cornell.edu/projects/kddcup/datasets.html} is a citation network of papers about high-energy physics. The edges represent citations between papers. 400 keywords extracted from the abstracts of all papers are set as the binary attributes indicating whether the keywords appear in the abstract of the paper or not. DBLP \footnote{http://dblp.uni-trier.de/} is a co-authorship network of computer science authors. The edges represent co-authorships. 21 conferences from five areas, i.e., database (\emph{EDBT}, \emph{ICDE}, \emph{PODS}, \emph{SIGMOD}, \emph{VLDB}), data mining (\emph{ICDM}, \emph{KDD}, \emph{PAKDD}, \emph{PKDD}, \emph{SDM}), information retrieval (\emph{CIKM}, \emph{ECIR}, \emph{SIGIR}), artificial intelligence (\emph{AAAI}, \emph{IJCAI}, \emph{NIPS}, \emph{UAI}), multimedia (\emph{ICMCS}, \emph{MIR}, \emph{MM}, \emph{SIGGRAPH}), are set as binary attributes indicating whether the author has published papers in the conferences or not. POLBLOG \cite{Perozzi2014Focused} is a hyperlink network of online blogs that discuss political issues. The edges represent hyperlinks between blogs. Attributes are keywords in their text. Dataset statistics and some experiment statistical results are given in Table 1 and Table 2, respectively.
\begin{table}
\renewcommand{\arraystretch}{1.3}
\caption{Dataset Statistics. $n$: node number, $m$: edge number, $r$: attribute number.}
\label{T1}
\centering
\begin{tabular}{|c|c|c|c|}
\hline
Dataset & $n$ & $m$ & $r$  \\
\hline
arXiv & 29555 &	352807 &	400  \\
\hline
DBLP & 78619 &	597591 &	21  \\
\hline
POLBLOG & 362 & 2233 & 44839  \\
\hline
\end{tabular}
\end{table}
\begin{table}
\renewcommand{\arraystretch}{1.3}
\caption{Experiment Statistical Results. \#: number of mined communities, $\overline{|C|}$: average size of mined communities, Time: Running time, `-NE' denotes the results are obtained by TSCM-NE on corresponding network.}
\label{T1}
\centering
\begin{tabular}{|c|c|c|c|}
\hline
Dataset & \# & $\overline{|C|}$ & Time (sec) \\
\hline
arXiv &	1 & 54.0 &	1.600e+1 \\
\hline
arXiv-NE &	1 & 44.0 &	1.049e+1 \\
\hline
DBLP &	5 & 721.2 &	9.225e+1 \\
\hline
DBLP-NE &	61 & 136.4 &	1.295e+2 \\
\hline
POLBLOG & 5 & 199.8  & 1.018e+2 \\
\hline
\end{tabular}
\end{table}

In arXiv, assuming a researcher has read two papers called ``super-poincare covariant quantization of the superstring'' and ``covariant quantization of the superstring''. He want to read more related papers. Two read papers are set as the sample nodes for TSCM. The target subspace mined by TSCM includes several focus attributes, such as \emph{string}, \emph{spin}, \emph{superstring}, \emph{quantization}, \emph{constraint}, \emph{operator} and \emph{covariant}, etc.. In this target subspace, one target community with 54 papers about quantization of the superstring is extracted, including papers called ``covariant quantization of superstrings without pure spinor constraints'', ``an introduction to the covariant quantization of superstrings'' and ``towards covariant quantization of the supermembrane'', etc.. The extracted target community can be recommended to the researcher. This case study shows that the TSCM is helpful for product recommendation. In order to evaluate the effect of the sample information extension technique in target subspace mining process, we also conduct the method TSCM-NE in arXiv, which removes the sample information extension procedure from TSCM by setting the provided two sample nodes as the exemplar nodes directly. In this case, the focus attributes in the mined target subspace are keywords that appear in abstracts of two provided papers simultaneously. Besides focus attributes that are in the target subspace of TSCM, many more focus attributes are included in the target subspace of TSCM-NE, such as \emph{curve}, \emph{mass}, \emph{time}, \emph{vertex} and \emph{dimension}, etc.. Most of these extra focus attributes have no tightly relations to each other. The target subspace of TSCM is more reasonable, since its focus attributes concentrate on the topic of quantization of the superstring more closely. One target community with 44 papers is detected by the TSCM-NE. The F1 score between the target community detected by TSCM and that detected by TSCM-NE is 0.75. It means the two detected communities are not very similar to each other. Thus the sample information extension procedure is essential for mining the more reasonable target subspace and target communities.

In DBLP, assuming we want to advertise a new journal to researchers. Suppose two authors, Charu Aggarwal and Yizhou Sun, have published papers in this new journal. They are set as the sample nodes for TSCM. The mined target subspace includes four focus attributes, i.e., \emph{ICDM}, \emph{SIGKDD}, \emph{PAKDD}, \emph{SDM}. They are mainly about data mining. This is reasonable because both Charu Aggarwal and Yizhou Sun are famous data mining experts. Five target communities are extracted in this subspace. We can advertise the journal to researchers in extracted target communities. This case study shows that the TSCM can be used for advertising. We also conduct the TSCM-NE in DBLP. Since there is only one conference \emph{PAKDD} that Charu Aggarwal and Yizhou Sun have published papers in simultaneously, the mined target subspace of TSCM-NE only contains one focus attribute \emph{PAKDD}. Some related attributes about data mining fail to be included as focus attributes without sample information extension procedure. 61 target communities are extracted by TSCM-NE. Most of them are very small with less than 10 nodes. The average size of target communities of TSCM-NE is 136 which is much less than that of TSCM in Table 1. The target communities of TSCM are communities about data mining, while those of TSCM-NE are only communities in \emph{PAKDD}. 

\begin{figure}
\centering
\includegraphics{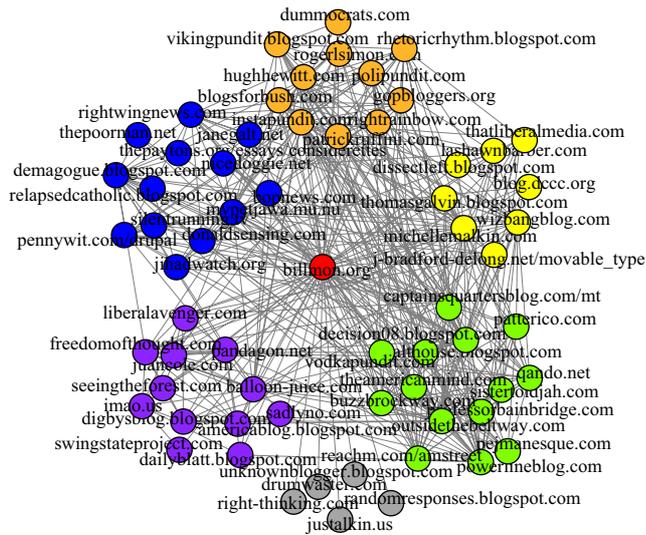}
\caption{The ego net of \textit{billmon.org}. Node sets in green, blue, purple, orange and yellow are five sets of exemplar nodes.}
\label{F6}
\end{figure}

Finally, we use TSCM to analyze attribute subspaces and communities around a famous blog ``billmon.org'' in POLBLOGS. Billmon is one of the leading bloggers writing pieces on domestic politics, Iraq war and US economy. We set ``billmon.org'' as the sample node. Its ego net is illustrated in Fig. 7. Five subspaces are mined from five neighborhood communities which are colored in green, blue, purple, orange and yellow, respectively. Every grey node forms a neighborhood community by itself, so there is no reasonable subspace computed from each of them. Since the dataset was created during the Iraq war in 2005, some common attributes exist in all five subspaces, such as \emph{anti}, \emph{Iraq} and \emph{war}, etc.. Besides common attributes, subspaces green and orange contain specific attributes such as \emph{Afghanistan}, \emph{terrorists}, \emph{bombers} etc., which indicate communities in them are interested in terrorism issues. Subspaces blue and purple have particular attributes such as \emph{business}, \emph{job}, \emph{economic} etc., which mean communities in them are interested in US economy. Subspace yellow has specific attributes such as \emph{freedom}, \emph{race}, \emph{justice} and \emph{religious} etc., which indicate community in it is interested in democratic politics. This case study shows that TSCM can be used to analyze subspaces and communities around an important node.

\section{Conclusion}
In this paper, we study an application oriented problem of mining the target subspace and the set of target communities for any specific application. Two provided sample nodes have limited information about the target subspace. A sample information extension method is proposed to use some neighbors of the sample nodes to help form the exemplar node set and the target subspace is computed from the set of exemplar nodes. The network is re-weighted based on the target subspace and a network backbone is constructed. The cohesive parts of the network backbone are set as the community seeds. Finally, the target communities are extracted by locally expanding the community seeds, and then redundant communities are eliminated. The extensive experiments on synthetic networks demonstrate the effectiveness and efficiency of TSCM, and the applications on real-world networks show its application values.

\begin{acknowledgements}
This work is supported by National Key Basic Research Program of China (2013CB329603), National Natural Science Foundation of China (U1636105).
\end{acknowledgements}

\bibliographystyle{spphys}       % APS-like style for physics
\bibliography{Manuscript}   % name your BibTeX data base

\begin{thebibliography}{10}
\providecommand{\url}[1]{{#1}}
\providecommand{\urlprefix}{URL }
\expandafter\ifx\csname urlstyle\endcsname\relax
  \providecommand{\doi}[1]{DOI \discretionary{}{}{}#1}\else
  \providecommand{\doi}{DOI \discretionary{}{}{}\begingroup
  \urlstyle{rm}\Url}\fi

\bibitem{Zhou2009Graph}
Y.~Zhou, H.~Cheng, J.X. Yu, Proc. VLDB \textbf{2}(1), 718 (2009).
\newblock \doi{10.14778/1687627.1687709}

\bibitem{Xu2012A}
Z.~Xu, Y.~Ke, Y.~Wang, H.~Cheng, J.~Cheng, in \emph{SIGMOD} (ACM, 2012), pp.
  505--516.
\newblock \doi{10.1145/2213836.2213894}

\bibitem{Ruan2013Efficient}
Y.~Ruan, D.~Fuhry, S.~Parthasarathy, in \emph{Proceedings of the 22nd
  international conference on world wide web} (IW3C2, 2013), pp. 1089--1098

\bibitem{Moser2009Mining}
F.~Moser, R.~Colak, A.~Rafiey, M.~Ester, in \emph{SDM}, vol.~9 (SIAM, 2009),
  vol.~9, pp. 593--604.
\newblock \doi{10.1137/1.9781611972795.51}

\bibitem{Silva2012Mining}
A.~Silva, W.~Meira~Jr, M.J. Zaki, Proc. VLDB \textbf{5}(5), 466 (2012).
\newblock \doi{10.14778/2140436.2140443}

\bibitem{Gunnemann2013Spectral}
S.~G{\"u}nnemann, I.~F{\"a}rber, S.~Raubach, T.~Seidl, in \emph{ICDM} (IEEE,
  2013), pp. 231--240.
\newblock \doi{10.1109/ICDM.2013.110}

\bibitem{Gunnemann2014Gamer}
S.~G{\"u}nnemann, I.~F{\"a}rber, B.~Boden, T.~Seidl, Knowledge and information
  systems \textbf{40}(2), 243 (2014).
\newblock \doi{10.1007/s10115-013-0640-z}

\bibitem{Galbrun2014Overlapping}
E.~Galbrun, A.~Gionis, N.~Tatti, Data Mining and Knowledge Discovery
  \textbf{28}(5-6), 1586 (2014).
\newblock \doi{10.1007/s10618-014-0373-y}

\bibitem{Huang2015Dense}
X.~Huang, H.~Cheng, J.X. Yu, Information Sciences \textbf{314}, 77 (2015).
\newblock \doi{10.1016/j.ins.2015.03.075}

\bibitem{Raghavan2007Near}
U.N. Raghavan, R.~Albert, S.~Kumara, Phys. Rev. E \textbf{76}(3), 036106
  (2007).
\newblock \doi{10.1103/PhysRevE.76.036106}

\bibitem{Wu2015Multi}
P.~Wu, L.~Pan, PLoS ONE \textbf{10}(5), e0126845 (2015).
\newblock \doi{10.1371/journal.pone.0126845}

\bibitem{Perozzi2014Focused}
B.~Perozzi, L.~Akoglu, P.~Iglesias~S{\'a}nchez, E.~M{\"u}ller, in \emph{SIGKDD}
  (ACM, 2014), pp. 1346--1355.
\newblock \doi{10.1145/2623330.2623682}

\bibitem{Mcauley2014Discovering}
J.~Mcauley, J.~Leskovec, ACM Transactions on Knowledge Discovery from Data
  (TKDD) \textbf{8}(1), 4 (2014).
\newblock \doi{10.1145/2556612}

\bibitem{Lancichinetti2009Detecting}
A.~Lancichinetti, S.~Fortunato, J.~Kert{\'e}sz, New Journal of Physics
  \textbf{11}(3), 033015 (2009).
\newblock \doi{doi:10.1088/1367-2630/11/3/033015}

\bibitem{Zhou2010Clustering}
Y.~Zhou, H.~Cheng, J.X. Yu, in \emph{ICDM} (IEEE, 2010), pp. 689--698.
\newblock \doi{10.1109/ICDM.2010.41}

\bibitem{Yang2013Community}
J.~Yang, J.~McAuley, J.~Leskovec, in \emph{ICDM} (IEEE, 2013), pp. 1151--1156.
\newblock \doi{10.1109/ICDM.2013.167}

\bibitem{Akoglu2012Pics}
L.~Akoglu, H.~Tong, B.~Meeder, C.~Faloutsos, in \emph{SDM} (SIAM, 2012), pp.
  439--450

\bibitem{Pool2014Description}
S.~Pool, F.~Bonchi, M.v. Leeuwen, ACM Transactions on Intelligent Systems and
  Technology (TIST) \textbf{5}(2), 28 (2014).
\newblock \doi{10.1145/2517088}

\bibitem{Lancichinetti2009Benchmarks}
A.~Lancichinetti, S.~Fortunato, Phys. Rev. E \textbf{80}(1), 016118 (2009).
\newblock \doi{10.1103/PhysRevE.80.016118}

\bibitem{Vincent2008Fast}
D.B. Vincent, G.~Jean-Loup, L.~Renaud, L.~Etienne, J. Stat. Mech.
  \textbf{2008}(10), P10008 (2008).
\newblock \doi{10.1088/1742-5468/2008/10/P10008}

\end{thebibliography}

\end{document}